\documentclass[10pt,letterpaper,twocolumn]{article}

\usepackage{ol2}
\usepackage[draft]{hyperref}
\usepackage{amsmath}
\usepackage{epstopdf}

\begin{document}

\twocolumn[ 
\title{A high-fidelity single-photon source\\ based on a type-II optical parametric oscillator}

\author{Olivier Morin,$^1$ Virginia D'Auria,$^2$ Claude Fabre,$^1$ and Julien Laurat$^{1,*}$}

\address{
$^1$Laboratoire Kastler Brossel, Universit\'{e}
Pierre et Marie Curie, Ecole Normale Sup\'{e}rieure, CNRS, \\4 place
Jussieu, 75252 Paris Cedex 05, France\\
$^2$Laboratoire de Physique de la Mati\`{e}re Condens\'{e}e, CNRS UMR 7336,
Universit\'{e} de Nice - Sophia Antipolis,\\
 Parc Valrose, 06108 Nice Cedex 2, France\\
$^*$Corresponding author: julien.laurat@upmc.fr
}

\begin{abstract}
Using a continuous-wave type-II optical parametric oscillator below threshold, we have demonstrated a novel source of heralded single-photons with high-fidelity. The generated state is characterized by homodyne detection and exhibits a 79$\%$ fidelity with a single-photon Fock state ($91\%$ after correction of detection loss). The low admixture of vacuum and the well-defined spatiotemporal mode are critical requirements for their subsequent use in quantum information processing. \
\end{abstract}

\ocis{270.0270, 270.6570, 270.5290,190.4970}
 ]


\noindent The reliable generation of single-photon states is a central resource for the development of quantum information sciences and technologies, including quantum communication and computing \cite{Buller2010,Sangouard2012}. For instance, since the seminal proposal by Knill, Laflamme and Milburn \cite{KLM}, single-photons are indeed at the heart of linear-optical quantum computation (LOQC) \cite{Kok2007}. Practical implementations however require to generate such states with a low admixture of vacuum as efficient LOQC protocols are constrained by loss thresholds \cite{Varnava2008}. For instance, the best known figure to date,  which applies to cluster state computation, is a $1/2$ overall loss tolerance \cite{Gong2010}, i.e. the product of the source fidelity and detector efficiency has to be above this value. This constraint puts a challenging demand on single-photon generation. Additionally, linear-optical processing cannot increase the fidelity of the state, even with multiple imperfect sources \cite{Berry2010}. Moreover, gate implementation requires to make single-photons interfere with a high visibility and thus to be emitted into a well-controlled spatiotemporal mode. 
 
\begin{figure}[b!]
\vspace{-0.2cm}
\centerline{\includegraphics[width=0.95\columnwidth]{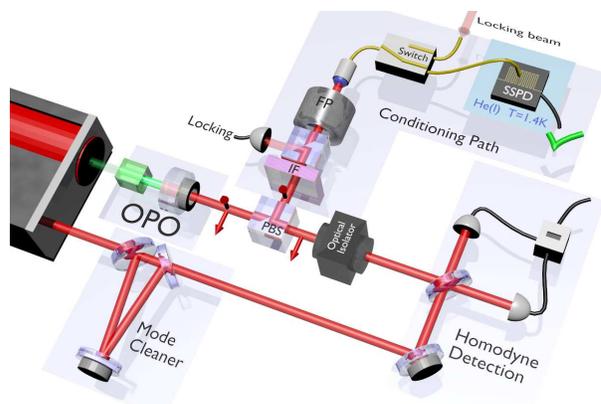}}
\caption{Experimental setup. A continuous-wave type-II optical parametric oscillator (OPO) is pumped far below threshold. The orthogonally-polarized signal and idler modes are separated by a polarizing beam-splitter (PBS). The idler mode is frequency filtered (Conditioning Path) and detected by a superconducting single-photon detector (SSPD). Given a detection event, the heralded single-photon is characterized by homodyne detection. FP: Fabry-Perot cavity, IF: interferential filter.}
\label{fig1}
\end{figure}
In this letter, we demonstrate the generation of high-fidelity single-photons based on a frequency-degenerate type-II optical parametric oscillator (OPO). Our heralded source combines the two important previous properties: a low admixture of vacuum and a very-well defined spatiotemporal mode due to the OPO cavity.

The generation technique consists in using photon pairs emitted in two distinct modes: the detection of a single-photon on one mode heralds the preparation of a single-photon into the other one \cite{Hong86}. This conditional preparation has been widely used so far with a variety of physical systems, including atomic cascade, atomic ensemble \cite{Laurat2006_Atoms} and pulsed single-pass parametric down-conversion (PDC) \cite{Lvovsky2001, Zavatta2004, Ourjoumtsev2006, Huisman2009}. More recently, continuous-wave OPOs, where PDC occurs in cavity, have also been considered for this purpose \cite{Nielsen2007}. 

In this case, two methods can be used to obtain the two distinct modes.  The first one is based on tapping a very small fraction of a squeezed vacuum \cite{Furusawa}, leading intrinsically to a low count rate. The other method consists in using non-degenerate modes emitted into the same spatial mode of the OPO cavity. Using a type-I non-linear interaction, signal and idler have the same polarization: the only degree of freedom is thus the frequency. The two correlated modes can be then two modes with frequencies separated by multiples of the cavity free spectral range, as demonstrated in \cite{Neergaard2007}. However, this configuration requires for the homodyne tomography to use
 local oscillators with shifted frequencies. The other possibility is to use a type-II interaction: the frequency-degenerate signal and idler modes have orthogonal polarizations and can be easily separated. This technique, which was not implemented so far, leads to a simpler apparatus and enables us to reach a higher fidelity.

The experimental setup is sketched on Fig. \ref{fig1}. A continuous-wave frequency-doubled Nd:YAG laser (Diabolo, Innolight) pumps below threshold a triply-resonant OPO made of a semimonolithic linear cavity \cite{Laurat2004}. The input mirror is directly coated on one face of a 10 mm-long KTP crystal (Raicol), with an intensity reflection of 95$\%$ for the 532 nm pump and high reflection at 1064 nm. The output coupler, with a 38 mm radius of curvature, is highly reflective for the pump and has a reflection of 90$\%$ for the infrared. The free spectral range is equal to $\Delta=4.3$ GHz. The internal losses at 1064 nm are estimated to be 0.4\%, giving a cavity bandwidth $\gamma=60$ MHz. The triple resonance condition is achieved by adjusting the frequency of the laser and the temperature of the crystal. The OPO cavity is locked on the pump resonance by the Pound-Drever-Hall technique. 

At the output of the OPO, the orthogonally-polarized signal and idler modes are spatially separated thanks to a polarizing beam splitter. The idler is directed to frequency filtering elements (conditioning path on Fig. \ref{fig1}) to remove non-degenerate modes. Indeed, the output of the OPO cavity contains a great number of modes separated by the free spectral range: each mode at frequency $\omega_0+p\Delta$ is entangled with the mode at $\omega_0-p\Delta$, where $p$ is an integer. In order to herald the generation of a single-photon at the central frequency $\omega_0$, it is thus necessary to reject all these non-degenerate modes. A first filtering stage is based on a commercial interferential filter (Barr Associates), with a bandwidth equal to 0.5 nm. The remaining modes are then filtered by a linear Fabry-Perot cavity. With a length equal to 0.45 mm and a finesse of 1000, this cavity has a free spectral range of 330 GHz and a bandwidth close to 320 MHz. The bandwidth is chosen in order to be larger than the one of the OPO and the spectral range to be larger than the one of the interferential filter. This two-stage filtering enables us to reach a -25 dB rejection (0.3$\%$) of the unwanted modes. 

The experiment is conducted on a cyclic fashion: the filtering cavity is locked during 40 ms, then the measurement period starts for 60 ms.The filtering cavity is locked on resonance by the Dither-and-Lock technique. For this purpose, a counter-propagating beam is injected via an optical switch (Thorlabs OSW12-980E). The beam is rejected at the entrance of the cavity via an optical circulator and detected. During the measurement period, the optical switch is inverted, the filtered mode is detected by a superconducting single-photon detector (SSPD, Scontel) working at cryogenic temperature. The quantum efficiency at 1064 nm reaches $7\%$. Notably, dark noise is negligible (below 1 Hz), an important feature as false events degrade the fidelity of the heralded state \cite{DAuria2011}.

\begin{figure}[t!]
\centerline{\includegraphics[width=0.85\columnwidth]{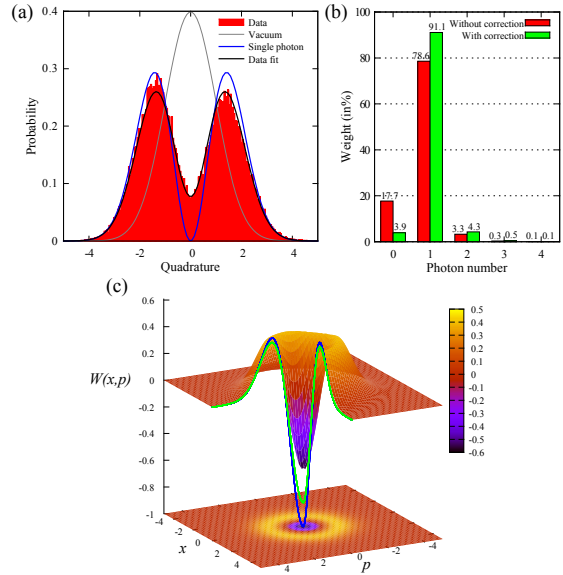}}
\caption{High-fidelity single-photon. (a) Measured quadrature obtained from 50000 acquisitions. The black solid line is a fit of the experimental data, while the blue solid line provides the distribution for a perfect single-photon state and the gray line for the vacuum.  (b) Diagonal elements of the density matrix, with and without correction from detection losses. (c) Corresponding Wigner function. The solid lines give the cross-section for a perfect single-photon and the experimental cross-section after correction. $x$ and $p$ denote quadrature components.}
\vspace{-0.4cm}\label{fig2}
\end{figure}

Given a detection event on the conditioning path, the heralded state is then characterized by quantum state tomography performed via a homodyne detection. The 6 mW local oscillator is provided by
the 1064 nm output of the laser, after spatial mode filtering by a high-finesse cavity (`mode cleaner' in Fig. \ref{fig1}). The homodyne detection (55 MHz bandwidth) is based on a pair of high quantum efficiency InGaAs photodiodes (Fermionics, 500 $\mu$m, ~97$\%$). An optical isolator just before the homodyne detection is required in our setup to remove light from the local isolator backscattered by the photodiodes. For each trigger event, the photocurrent is recorded with an oscilloscope (Lecroy Wavepro 7300A)  at  a sampling rate of 5 GS/s during 100 ns. The local oscillator phase is swept during the measurement. 

As the local oscillator is continuous, post-processing is required to extract the conditional state in the temporal mode $f(t)$.  This mode depends on the bandwidth of the OPO and of the filtering path. In our case, the filtering path is much wider and does not affect the temporal mode. The theoretical expression of the optimal mode is thus $f(t)=\sqrt{\pi\gamma}e^{-\pi\gamma|t|}$ \cite{Nielsen2007}. Overall, each trigger event leads to a single outcome of the quadrature measurement, which is obtained as $x=\int f(t)x(t)dt$, where $x(t)$ is the measured difference photocurrent signal from the homodyne detection. Accumulated measurements are then processed with a maximum likelihood algorithm \cite{Lvovsky2009}, which provides the density matrix of the heralded state and the corresponding Wigner function.  The value of $\gamma$ is optimized to obtain the highest contribution of single-photon, i.e. the largest negativity of the Wigner function. The resulting value (65 MHz) is slightly different from the OPO bandwidth as the gain of the electronics is not flat over the whole spectrum.


Results are displayed on Fig. \ref{fig2}. Figure \ref{fig2}(a) gives the histogram of the measured quadrature values, while fig. \ref{fig2}(b) and \ref{fig2}(c) show respectively the diagonal elements of the reconstructed density matrix and the corresponding Wigner function. The state is in good approximation a mixture of vacuum and single-photon state. The two-photon component is around 3$\%$, resulting from multiphoton pairs creating by the down-conversion process. The OPO is operated far below threshold (1 mW for a threshold of 80 mW) to limit this contribution. Without any corrections, the single-photon component reaches $78.6\pm0.5\%$ (Error bar is estimated following the method in \cite{Morin12}). To the best of our knowledge, this is the highest value reported to date. By taking into account the detection losses, we infer a value as high as $91\%$.  

Table \ref{table} gives the various efficiencies affecting the final result. Two contributions have to be distinguished. A first one comes from the overall losses coming from propagation and detection. This includes the propagation efficiency $\eta_\text{prop}$, the mode overlap limited by the  visibility $V$ of the local oscillator-signal interference $\eta_\text{vis}=V^2$, the efficiency of the photodiodes $\eta_\text{phot}$ and the electronic noise of the detection $\eta_\text{noise}$ (the electronic noise is 20 dB below vacuum noise at the central frequency) \cite{Kumar2011}. They sum up to an overall detection loss of $1-\eta_\text{tot}=$15$\%$, which is taken into account to give the values with correction. The second contribution is more fundamental and cannot be corrected as it is associated with the generation process. It depends on the OPO escape efficiency, given by $\eta_\text{OPO}=T/(T+L)$, where $T$ is the transmission of the output coupler and $L$ the intracavity loss. This value is estimated to be $\eta_\text{OPO}=96\%$. Given these estimated parameters, the expected vacuum component is 18\%, in very good agreement with the measured value.

\vspace{-0.4cm}\begin{table}
 \centering
  \caption{Experimental efficiencies}\begin{tabular}{cccc|c||c} \\ \hline
       $\eta_\text{noise}$ & $\eta_\text{phot}$ & $\eta_\text{vis}$ & $\eta_\text{prop}$ &  $\eta_\text{tot}$ & $\eta_\text{OPO}$ \\ \hline
     96\% & 97\% & (98\%)$^2$ & 95\% &85\% & 96\%\\ \hline
  \end{tabular}
  \label{table}
\end{table}

The heralding rate for single-photon generation is 30 kHz. Given the bandwidth of the OPO, it corresponds to a brightness of 400 photons/s per MHz. This rate is mainly limited by the losses in the conditioning path, which reaches 97$\%$: the quantum efficiency of the SSPD is 7$\%$ and the overall transmission (including optical switch and filtering elements) is 40$\%$. Corrected from the losses in this path, the rate can be close to 750 kHz. 

In conclusion, we have reported the generation of high-fidelity heralded single-photons, using a configuration based on a  type-II optical parametric oscillator below threshold. A 79\% fidelity has been demonstrated, mainly limited by the losses in the detection path. Thanks to the OPO cavity, the spatial mode enables to reach high interference visibilities without the need of additional filterings. Moreover, the frequency-degenerate interaction makes the operation much simpler than previous realizations. This practical tool can facilitate the implementation of various new experiments in quantum information processing \cite{Morin12}. The technique developed in \cite{Nielsen09} could be extended to our efficient scheme if time-gating of the single-photons is required.\\

This work is supported by the ERA-NET CHIST-ERA (QScale) and the BQR from UPMC. V. D'Auria acknowledges the support from the EC (Marie Curie program). C. Fabre and J. Laurat are members of the IUF.

\end{document}